\documentclass{PoS}

\usepackage{graphicx,amssymb,amsbsy,amsfonts,amssymb,amsmath}
\usepackage{tikz}
\usetikzlibrary{calc}

\title{Five-Point Two-Loop Amplitudes from Numerical Unitarity}

\ShortTitle{Multi-Loop Numerical Unitarity}

\author{Samuel Abreu,$^a$ Fernando Febres Cordero\footnote{Speaker.}\ ,$^a$
Harald Ita,$^a$ Ben Page\footnote{Speaker.}\ ,$^a$
        and Mao~Zeng\!~$^b$\\
        \llap{$^a$}Physikalisches Institut, Albert-Ludwigs-Universit\"at, Freiburg\\
        D-79104 Freibug, Germany\\
        \llap{$^b$}Mani L.~Bhaumik Institute for Theoretical Physics \\
        UCLA Department of Physics and Astronomy \\
        Los Angeles, CA 90095, USA\\
        E-mail: \email{ffebres@physik.uni-freiburg.de}, \email{ben.page@physik.uni-freiburg.de}}

\abstract{We present advances in the development of the numerical unitarity
method for the computation of multi-loop amplitudes in QCD. As an
application, we show results for all the leading-color two-loop
five-gluon helicity amplitudes. The amplitudes are reduced to a linear
combination of master
integrals by employing unitarity-compatible integration-by-parts identities, and
the corresponding integral coefficients are computed in an exact manner on
rational phase-space points through finite fields arithmetics.}

\FullConference{Loops and Legs in Quantum Field Theory (LL2018)\\
		29 April 2018 - 04 May 2018\\
		St. Goar, Germany}

\begin{document}
\speaker{Fernando Febres Cordero and Ben Page}

\section{Introduction}
In the decade to come the LHC will be testing the Standard Model of particle
physics (SM) to an unprecedented level of accuracy. This provides a
natural pressure for innovation within the theory community, as
theoretical uncertainties associated to a large amount of observables
will need to be reduced to the few-percent level. To achieve this,
both next-to-next-to-leading-order (NNLO) QCD corrections and 
next-to-leading-order (NLO) electroweak corrections will need to be
computed for processes involving many scales. 
Though our capacity to perform the latter type of corrections has matured over
the last few years, the former stills present a variety of challenges.
NNLO QCD corrections are nevertheless highly 
desirable---relevant processes 
include $3$-jet, $H+2$-jet, $V+2$-jet, $t\bar t+$jet production (for
an extended discussion see ref.~\cite{Bendavid:2018nar}). 
For example, the importance of NNLO QCD corrections to $W$ production with two
$b$ jets has been recently
highlighted in ref.~\cite{Anger:2017glm} where an analysis based on
exclusive sums shows that the ${\cal O}(\alpha_s^2)$ corrections can be
sizable and necessary to stabilize the perturbative series of
associated observables, which are relevant for ongoing
and future studies of the $b$-quark Yukawa coupling (see for example
ref.~\cite{Aaboud:2017xsd,Sirunyan:2017elk}).

NNLO QCD corrections require many ingredients. Among the most
complicated is constructing a procedure to handle the different
IR-divergent contributions in such a way that the multiple phase-space
integrations required can be achieved numerically. Over the last 15
years a number of IR subtraction schemes have been developed,
which explicitly
manifest the cancellation of IR divergences and are starting to show
great potential for automation. Among them are antenna
subtraction~\cite{GehrmannDeRidder:2005cm,Currie:2013vh}, sector
improved residue subtraction~\cite{Czakon:2010td}, $q_T$
slicing~\cite{Catani:2007vq} and $N$-jettiness
slicing~\cite{Boughezal:2015eha,Gaunt:2015pea}.

Another challenging ingredient
which is the focus of these talks, 
is the
computation of two-loop amplitudes with a higher number of scales.
The first calculation of a two-loop five-gluon amplitude appeared five years
ago, for the special case of all positive helicities~\cite{Badger:2013gxa}. Since
then, a number of advances have been made for this particular
amplitude~\cite{Badger:2015lda,Gehrmann:2015bfy,Dunbar:2016aux},
followed by further calculations of the same helicity configuration at higher
multiplicity~\cite{Dunbar:2016gjb,Dunbar:2017nfy}. More recently the
complete set of helicity configurations were also studied with a
combination of numeric and analytic techniques~\cite{Badger:2017jhb},
and a follow up including a full reduction to master integrals was
presented at this conference.\footnote{See the talk by S. Badger in
these proceedings~\pos{PoS(LL2018)006}.} 
We also highlight the recent appearance of advanced
integration-by-parts reductions for massless two-loop five-point
amplitudes of refs.~\cite{Boels:2018nrr,Chawdhry:2018awn}.
Related promising
techniques for handling multi-scale two-loop problems have been
presented at this conference.\footnote{See the talks of W. Torres
Bobadilla \pos{PoS(LL2018)036} and K. Larsen \pos{PoS(LL2018)064}.}

In principle, these complex amplitude calculations can be performed by
constructing the full set of associated two-loop Feynman
diagrams. However, as is well known, the complexity of
computing
high-multiplicity processes in this way leads to very large
intermediate expressions
in analytic implementations (even
at one loop).
In this talk, we
focus on an approach~\cite{Abreu:2017xsl,Abreu:2017hqn} that exploits
the unitarity
method~\cite{Bern:1994zx,Bern:1994cg,Bern:1997sc,Britto:2004nc} in
order to numerically compute multi-loop amplitudes directly from
gauge-invariant building blocks.
The method makes use of an advanced decomposition of the amplitude's integrand
based on \textit{master integrands} and \textit{surface terms},
the latter being
constructed from unitarity-compatible integration-by-parts
identities. With the master integrals available,\footnote{See the
talks on these proceedings by J. Henn~\pos{PoS(LL2018)014} and
C. Papadopoulos~\pos{PoS(LL2018)015}.} we can then obtain the desired
amplitudes. As an application, we show a computation of the
leading-color two-loop five-gluon helicity amplitudes.

The remainder of these proceedings is divided into three sections. In
section~\ref{sec:uni}
we describe details of our computational framework. Section~\ref{sec:5g} shows
our results for the five-gluon helicity amplitudes, and finally we conclude in
section~\ref{sec:outlook} with an outlook about the prospects of the developed
techniques.

\section{Numerical unitarity}\label{sec:uni}

\begin{figure}[ht]
\begin{tikzpicture}[scale=2.0]
    \draw[dashed] (5.3,2.7) -- (5.3,0.22);
    \node at (1,3.4){\includegraphics[scale=0.22]{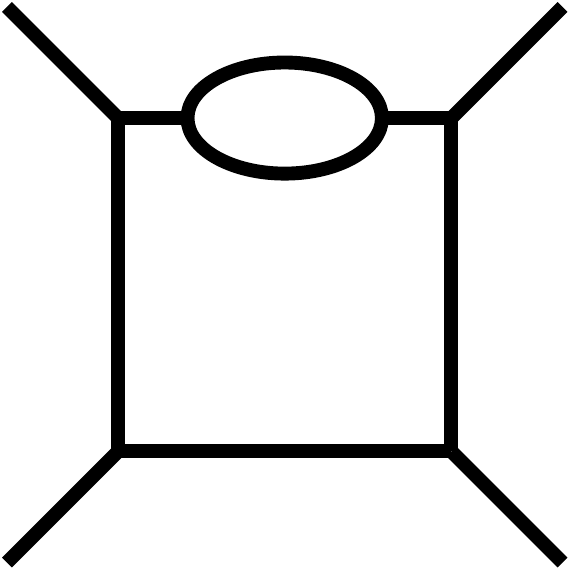}};
    \node at
    (2.55,3.4){\includegraphics[scale=0.22]{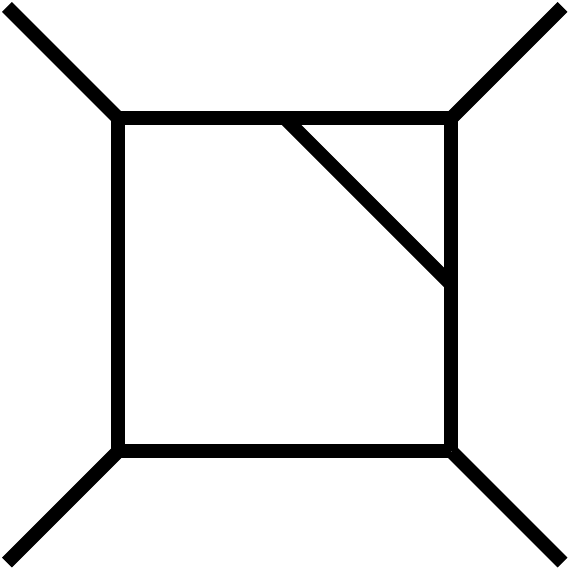}}; \node
    at (4.1,3.4){\includegraphics[scale=0.22]{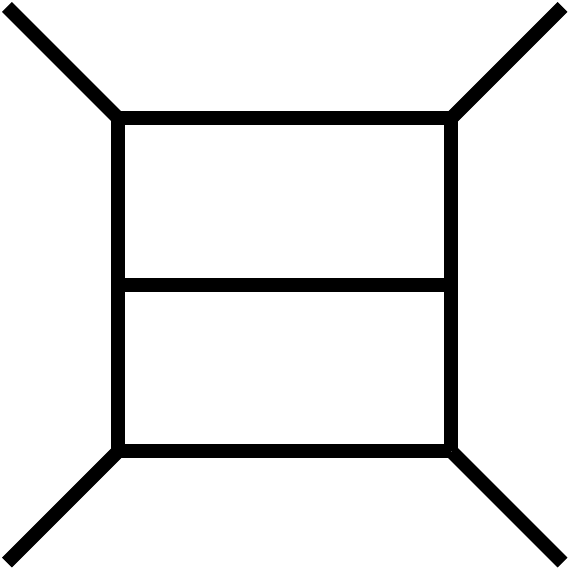}};
    \node at
    (0.55,2.4){\includegraphics[scale=0.22]{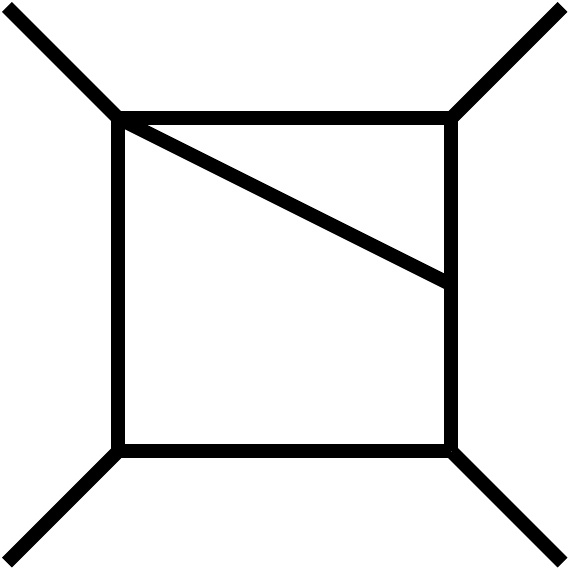}}; \node
    at (1.55,2.4){\includegraphics[scale=0.22]{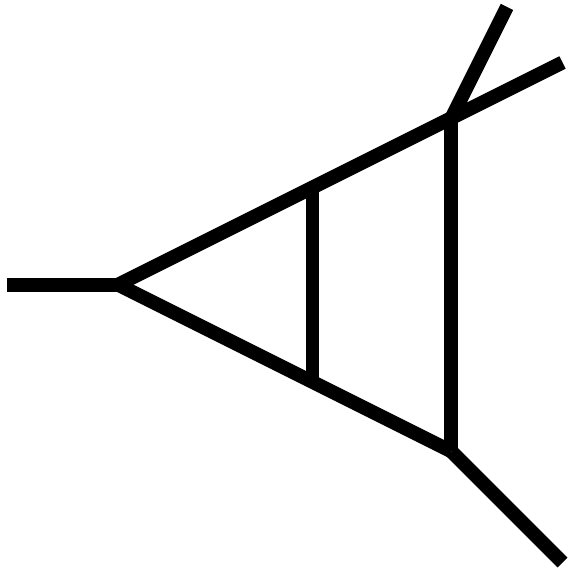}};
    \node at
    (2.55,2.4){\includegraphics[scale=0.22]{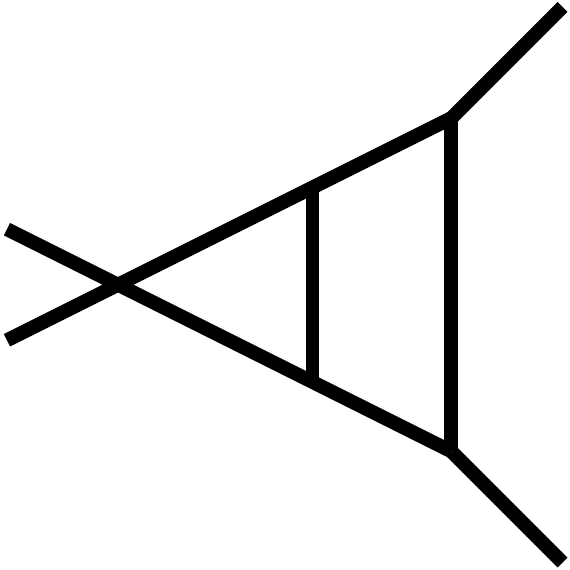}};
    \node at
    (3.55,2.4){\includegraphics[scale=0.22]{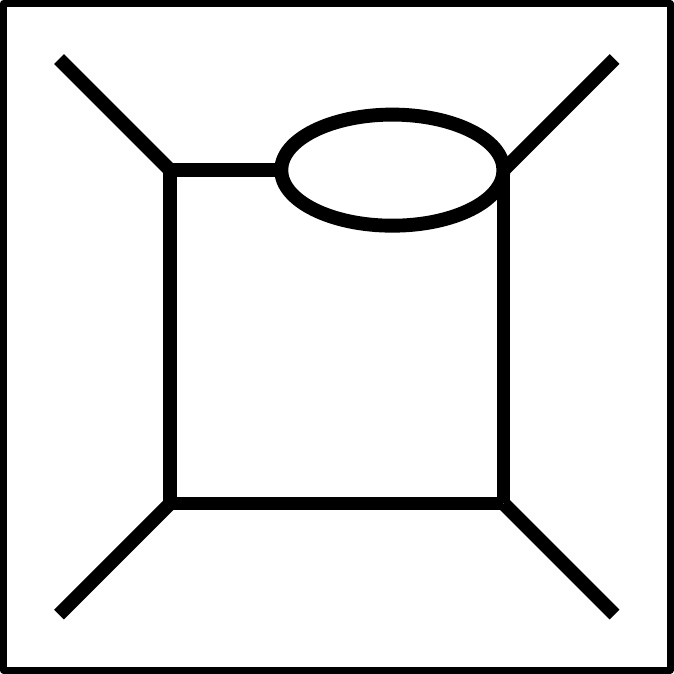}};
    \node at
    (4.55,2.4){\includegraphics[scale=0.22]{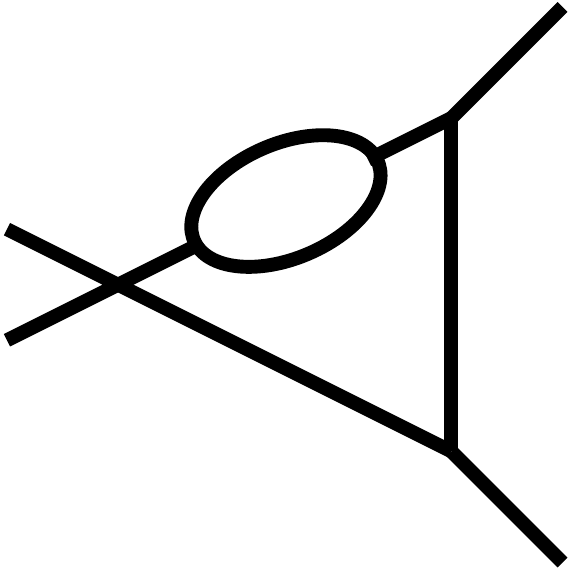}};
    \node at
    (5.8,2.4){\includegraphics[scale=0.22]{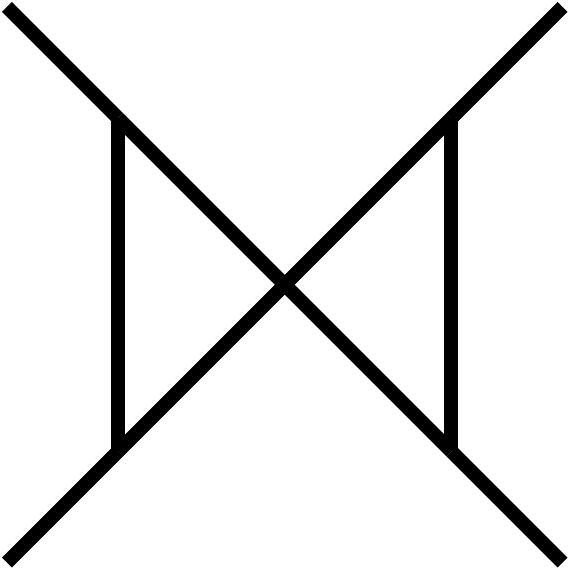}};
    \node at
    (6.5,2.4){\includegraphics[scale=0.22]{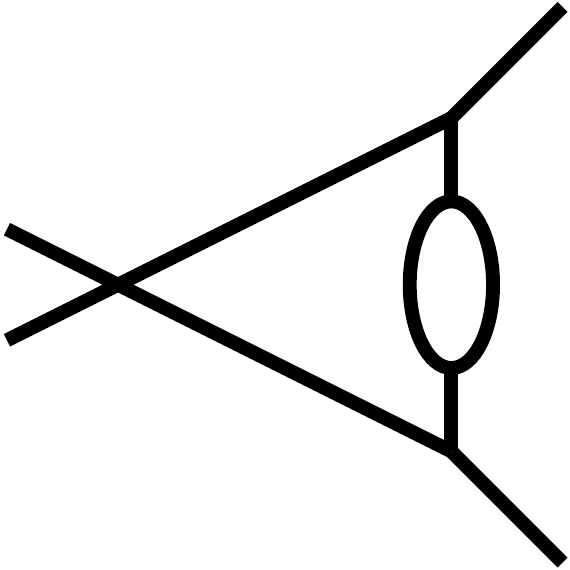}};
    \node at
    (0,1.4){\includegraphics[scale=0.22]{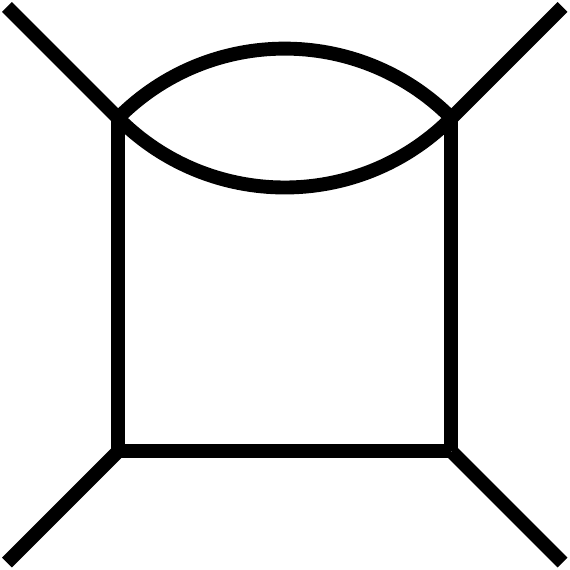}}; \node
    at
    (.8,1.4){\includegraphics[scale=0.22]{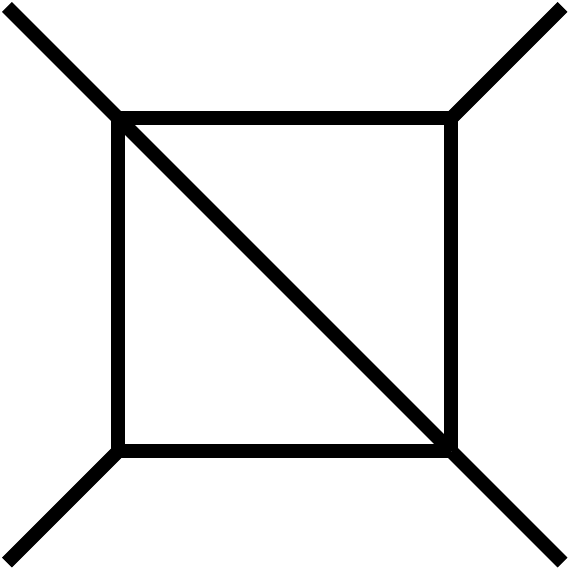}};
    \node at
    (1.6,1.4){\includegraphics[scale=0.22]{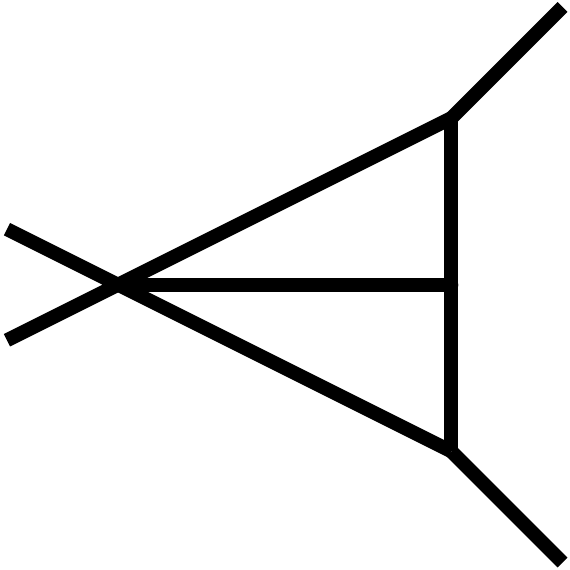}};
    \node at
    (2.4,1.4){\includegraphics[scale=0.22]{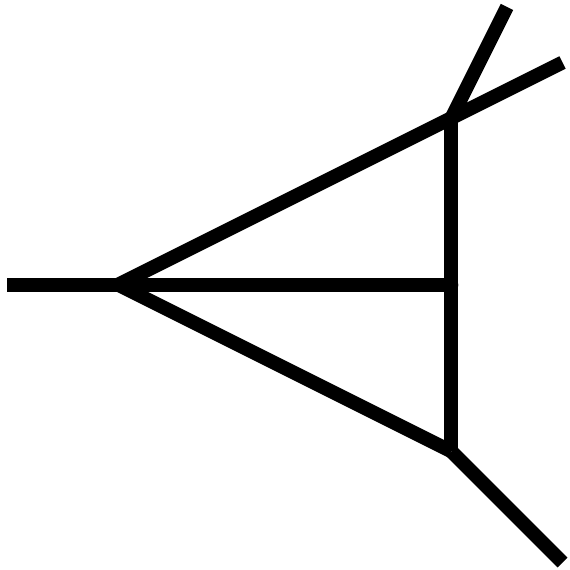}};
    \node at
    (3.2,1.4){\includegraphics[scale=0.22]{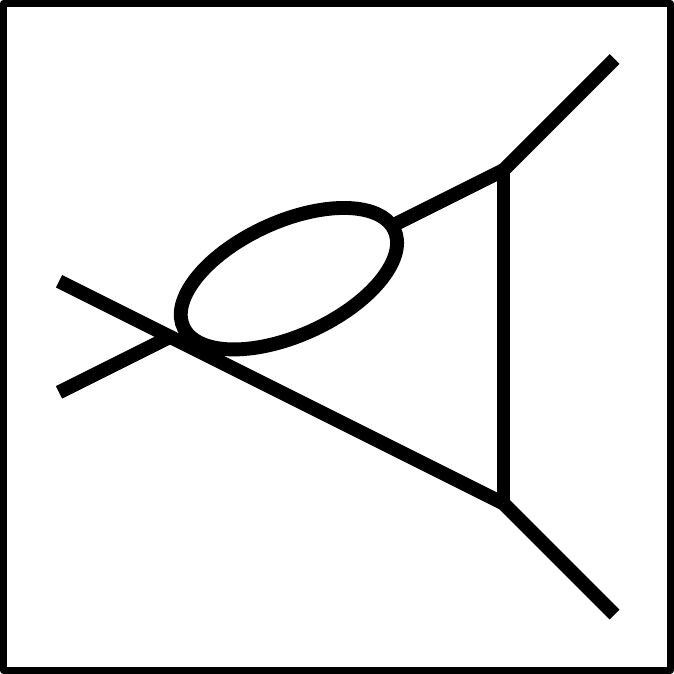}};
    \node at
    (4,1.4){\includegraphics[scale=0.22]{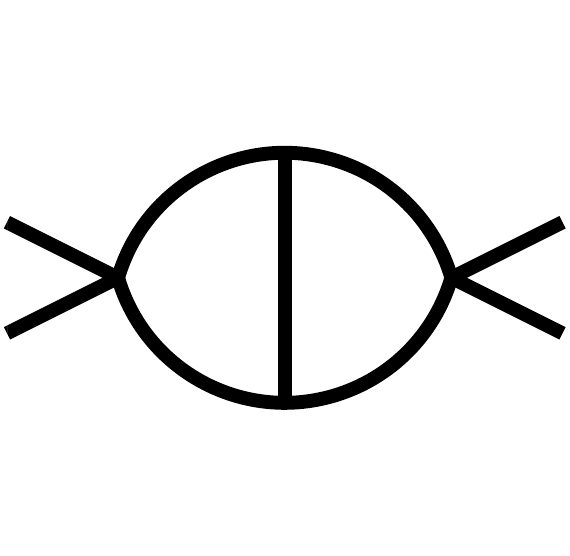}};
    \node at (4.8,1.4){\includegraphics[scale=0.22]{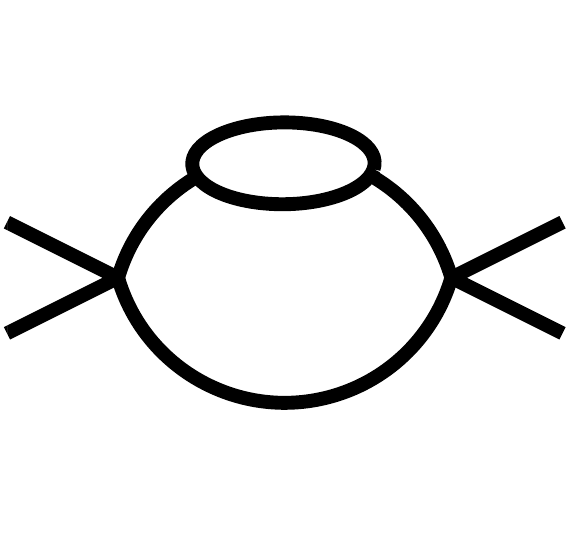}};
    \node at (5.8,1.4){\includegraphics[scale=0.22,angle=180]{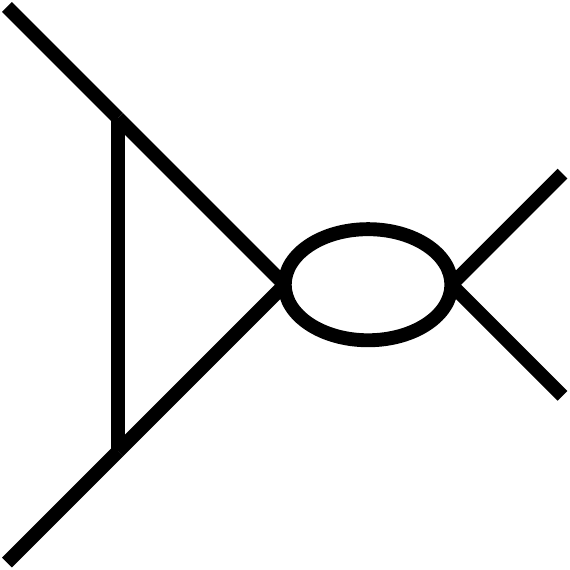}};
    \node at
    (6.5,1.4){\includegraphics[scale=0.22]{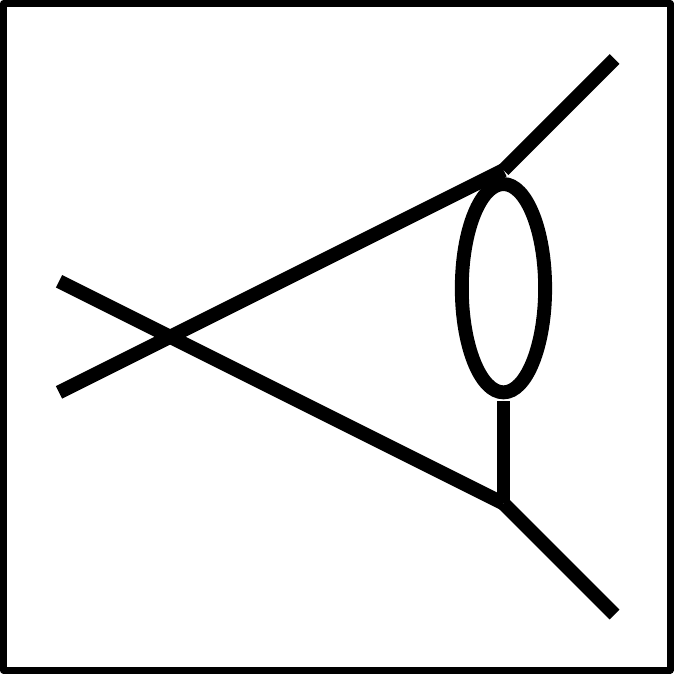}}; 
    \node at
    (1.75,.5){\includegraphics[scale=0.22]{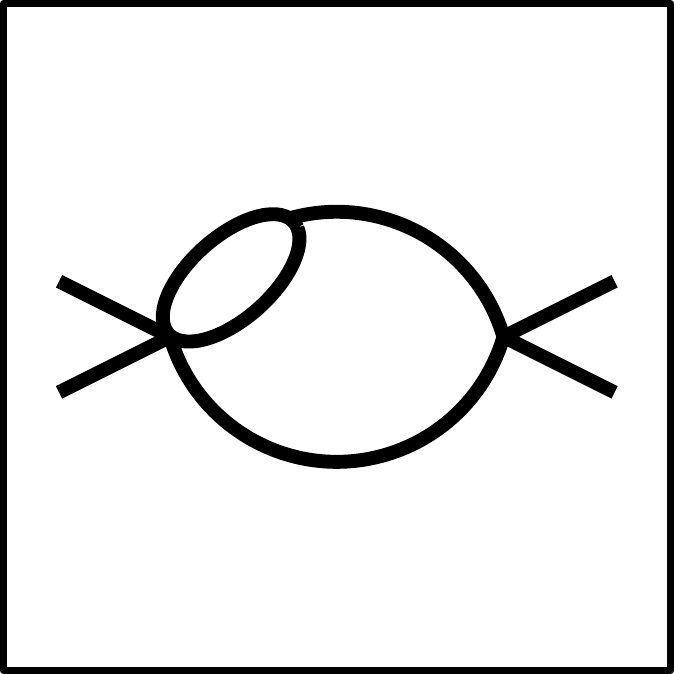}};
    \node at
    (3.35,.5){\includegraphics[scale=0.22]{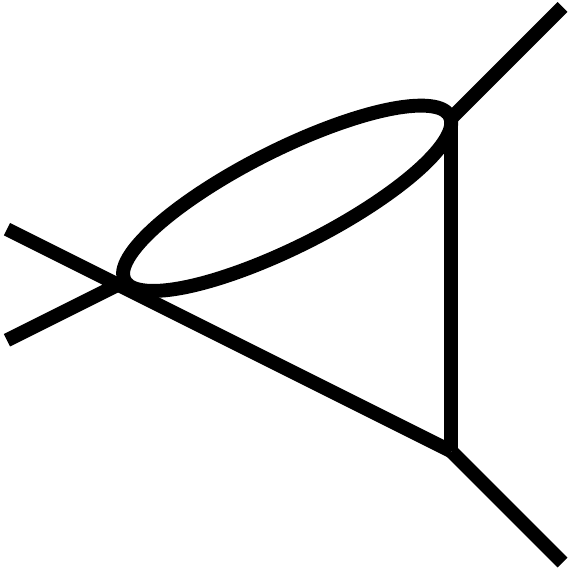}}; 
    \node at
    (5.8,.5){\includegraphics[scale=0.22]{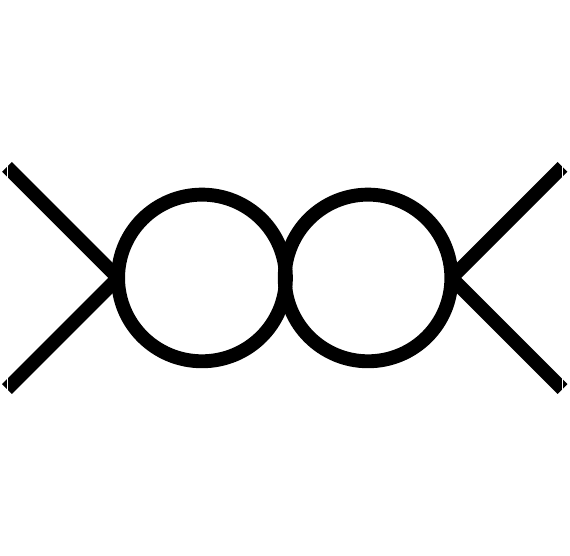}};
    \node at
    (6.5,.5){\includegraphics[scale=0.22]{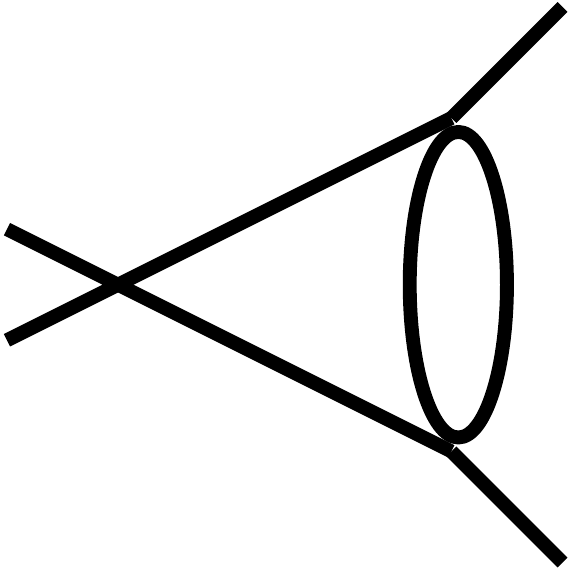}}; 
    \node at (2.55,-0.2){\includegraphics[scale=0.27]{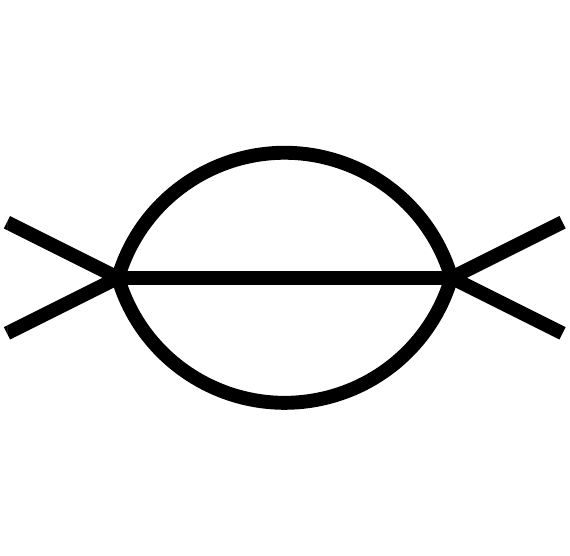}};
\end{tikzpicture} \caption{The planar hierarchy of propagator structures in a 2
$\to$ 2 massless amplitude. All edges represent massless propagators. We include
topologically inequivalent diagrams only. We highlight with boxes the diagrams that
do not have associated products of trees. To the left of the dashed line we collect
the sub-hierarchy of propagator structures that have the sunrise
as
their minimal cut.}
\label{fig:diagSunrise}
\end{figure}

In order to perform an amplitude calculation with numerical unitarity
we begin by classifying all propagator structures that can appear
in the integrand of the amplitude. These structures are represented by
a set of diagrams $\Delta=\{\Gamma_i\}$, which naturally can be
organized in a hierarchical way. As an
illustrative example, we show in figure~\ref{fig:diagSunrise} the full
hierarchy of propagator structures that appear for planar two-loop
four-point massless amplitudes. The top row contains all maximal
diagrams, those with seven propagators, and we go down to the bottom
of the hierarchy to the three-propagator diagram, the so called
\textit{sunrise} diagram.

To compute the amplitude, we start from the standard decomposition of an
amplitude in terms of master integrals:
\begin{equation}\label{eq:A}
\mathcal{A}^{(L)}=\sum_{\Gamma\in \Delta} \sum_{i\,\in\, M_\Gamma} c_{\Gamma,i} \,
{\mathcal I}_{\Gamma,i}\,,
\end{equation}
where $L$ is the number of loops, and $M_\Gamma$ represents the set of
master integrals associated to $\Gamma$ (notice that it can be empty
for many of the diagrams in $\Delta$). The process-dependent
coefficients $c_{\Gamma,i}$ are to be determined. We note that
they depend
on the dimensional regularization parameters $D$ and $D_s$.

Following a similar approach to the one-loop numerical unitarity
method~\cite{Ossola:2006us,Ellis:2007br,Giele:2008ve,Berger:2008sj}, we promote
the previous ansatz to the integrand level. That is, denoting the $L$-loop
integrand by $\mathcal{A}^{(L)}(\ell_l)$, with $\ell_l=\ell_1,\dots,\ell_L$, we
write:
\begin{equation}\label{eq:AL}
{\cal A}^{(k)}(\ell_l)=\,\sum_{\Gamma \in \Delta}\,\,\sum_{i\,\in\, M_\Gamma\cup
S_\Gamma} \frac{ c_{\Gamma,i} \,m_{\Gamma,i}(\ell_l)}{\prod_{j\in P_\Gamma}
\rho_j}\,, 
\end{equation}
where $\rho_j$ are the inverse propagators and $P_\Gamma$ represents
the set of propagators in diagram $\Gamma$. The $m_{\Gamma,i}$
functions form a basis parametrizing the integrand associated to the
diagram $\Gamma$ (up to corresponding power counting). In order for
the coefficients of the integrand level masters to correspond to those
in eq.~(\ref{eq:A}), the additional terms (with $i \in S_\Gamma$)
must integrate to zero, and hence we call them surface terms. We stress that the
$m_{\Gamma,i}(\ell_l)$ functions depend only on the kinematics of
the process, and not on the spin or helicity of the scattering
states.

The ansatz~(\ref{eq:AL}) holds for arbitrary loop momenta $\ell_l$, and
one can exploit this to construct linear systems of equations from
which one can compute all the coefficients $c_{\Gamma,i}$. In
generalized unitarity calculations, compact systems of equations can be
constructed by taking residues around the multi-propagator poles
corresponding to each diagram $\Gamma$,
\begin{equation}
       \sum_{\rm states}\prod_{i\in T_\Gamma} {\cal A}^{\rm tree}_i(\ell_l^\Gamma) =
\sum_{\substack{\Gamma' \ge \Gamma ,\\ i\,\in\,M_{\Gamma'}\cup S_{\Gamma'}}
} \frac{ c_{\Gamma',i}\,m_{\Gamma',i}(\ell_l^\Gamma)}{\prod_{ j \in P_j } \rho_j}\,. 
    \label{eq:factorizedAnsatz}
\end{equation}
On such a residue, the integrand factorizes into a product of
trees. The relations obtained from 
eq.~(\ref{eq:factorizedAnsatz}) are called 
\textit{cut equations}. We note that although starting at two loops
certain diagrams 
do not have a corresponding product of trees
(e.g.~the boxed ones in figure~\ref{fig:diagSunrise}),
following the procedure described in
ref.~\cite{Abreu:2017idw}
one can always find
enough cut equations to solve for all coefficients.

\subsection*{Master-surface decompositions}
The key component in writing down the integrand ansatz~(\ref{eq:AL}) is a
complete set of surface terms for each set of
propagator structures $\Gamma$. Precisely, one needs to construct the
full set of $m_{\Gamma,i}$ which satisfy
\begin{equation}
  \label{eq:SurfaceTerm}
  0 = \int [d^D\ell_l] \frac{m_{\Gamma, i}}
  {\prod_{j \in P_\Gamma} \rho_j}\,,
  \qquad\quad
  \textrm{with}\quad
  [d^D\ell_l]\equiv
  \prod_{l=1}^L \mathrm{d}^D \ell_l\,,
\end{equation}
where the $m_{\Gamma,i}$ are naturally polynomials in the loop
momenta.
The construction of such a set of surface terms follows the
description in~\cite{Abreu:2017xsl,Abreu:2017hqn,Ita:2015tya}. We
start by writing the integration-by-parts (IBP) relation,
\begin{equation}\label{eq:ibp-generic}
    \int [d^D\ell_l] \sum_k\frac{\partial}{\partial \ell^\nu_k}
    \left[\frac{u^\nu_k}{\prod_{j\in P_\Gamma}\rho_j}\right]=0.
\end{equation}
We note that the differential operator in 
eq.~\eqref{eq:ibp-generic} will in general generate terms with 
higher propagator power.
In order to account for this mismatch between eqs.~(
\ref{eq:SurfaceTerm})
and (\ref{eq:ibp-generic}) we choose vectors $u_k^\nu$ that preserve
the propagator structure associated to $\Gamma$. This is ensured by
defining the so called IBP-generating vectors~\cite{Gluza:2010ws}
according to:
\begin{equation} \label{eq:GKK}
u_k^\nu \frac{\partial}{\partial \ell_k^\nu} \rho_j = f_j \rho_j, \quad \forall j \in P_\Gamma,
\end{equation}
where $1\leq k\leq L$ and there is no sum over $j$. We require the $f_j$
functions to be polynomials in the loop momentum components,
which we parametrize with a convenient set of variables.

To find the necessary IBP-generating vectors, we start from the
ansatz
\begin{equation} u_k^\nu \frac{\partial}{\partial \ell_k^\nu}= \left( u^{\rm
loop}_{ka} \ell_a^\nu +  u^{\rm ext}_{kc} p_c^\nu \right)
\frac{\partial}{\partial \ell_k^\nu}, \end{equation}
where the labels $a$ and $c$ are summed over. Then eq.~(\ref{eq:GKK}) becomes~\cite{Abreu:2017hqn}:
\begin{equation} \label{eq:moduleSyz}
\left( u^{\rm loop}_{ka} \ell_a^\nu +  u^{\rm ext}_{kc} p_c^\nu \right)
\frac{\partial}{\partial \ell_k^\nu} 
\begin{pmatrix} \rho_{j(1)} \\ \rho_{j(2)} \\ \vdots \\ \rho_{j(|\Gamma|)}
\end{pmatrix} - 
\begin{pmatrix} f_{j(1)} \rho_{j(1)} \\ f_{j(2)} \rho_{j(2)} \\ \vdots \\
f_{j(|\Gamma|)} \rho_{j(|\Gamma|)} \end{pmatrix}
=0\, .
\end{equation}
This is a polynomial relation, known as a \emph{syzygy}
equation, which can be solved using algorithms
from computational algebraic geometry. We use the software
\texttt{Singular}~\cite{DGPS} to find solutions for all required
diagrams in our calculations. Analytic solutions can be found in under
a second for the most complicated case necessary for the results in
section~\ref{sec:5g}.

As a simple example of a surface term let us consider this procedure in the case of the one-loop
one-mass triangle diagram with propagators and external kinematics given by
\begin{align}\begin{split}
  \rho_1 = &(\ell + p_1)^2, \quad \quad \rho_2 = \ell^2,
  \quad\quad \rho_3 = (\ell-p_2)^2,\\
  & \quad  p_1^2 = p_2^2= 0, \quad \quad (p_1+p_2)^2 = s.
\end{split}\end{align}
We find that there is a single generating vector,
\begin{align}\begin{split}
  u^\nu &= u^{\mathrm{ext}}_{1} p_1^\nu + u^{\mathrm{ext}}_{2} p_2^\nu + u^{\mathrm{loop}} \ell^\nu\\
  \label{eq:TriangleGenerator}
  u^\nu &= (\rho_3 - \rho_2) p_1^\nu + (\rho_1 + \rho_2) p_2^\nu
          + (-s + 2 \rho_3 - 2 \rho_2) \ell^\nu.
\end{split}\end{align}
As this satisfies eq.~(\ref{eq:GKK}), the result of inserting this
expression into eq.~(\ref{eq:ibp-generic}) produces a result with no raised propagator powers,
\begin{equation}
  0 = \int \mathrm{d}^D\ell \frac{\partial}{\partial l^\nu} \frac{u^\nu}{\rho_1 \rho_2 \rho_3}
    = \int \mathrm{d}^D\ell \frac{1}{\rho_1 \rho_2 \rho_3}\left[ - (D-4) s - 2(D-3) \rho_2 + 2(D-3)\rho_3 \right].
\end{equation}
This relation can then be used either as a surface term,
or solved to obtain the well known IBP relation that
relates the scalar triangle to the scalar bubble. 
Techniques for
analytically solving a system of surface terms
to generate IBP relations are explored
in~\cite{Kosower:2018obg}.

Once a set of IBP-generating vectors is produced for a given diagram, a full
decomposition $\{m_{\Gamma,i}\}$ is achieved by multiplying them with
 a complete set of irreducible numerators respecting power
 counting~\cite{Ita:2015tya}, up until
the dimension of the space $S_\Gamma$ is filled. The complement of the
space can then be filled with independent insertions to produce the
corresponding master integrands in $M_\Gamma$. The full decomposition
of all planar five-point massless amplitudes in {\tt C++} code
takes about 3.5MB of
data, and we expect it to be possible to construct even more compact
expressions in the future.

\subsection*{Numerical unitarity and finite fields}

In recent years it has been demonstrated that so-called ``finite
fields'' can aid in the calculation of multi-loop scattering
amplitudes \cite{vonManteuffel:2014ixa,Peraro:2016wsq}. By exploiting
their mathematical structure, one can use computer integers to
obtain exact results in calculations involving only field operations.
In contrast to more commonly seen floating-point numerics, it is not
possible to solve arbitrary polynomial equations and this presents
challenges for constructing a numerical unitarity approach, many of
which were solved in \cite{Peraro:2016wsq}.

In the numerical unitarity method it is necessary to use
loop momenta satisfying the on-shell conditions associated to each of
the propagator structures $\Gamma$. As these are a system of
simultaneous quadratic equations this is a non-trivial exercise when
employing finite fields. To parametrize these ``on-shell phase
spaces'' we employ an adaptation of the van Neerven-Vermaseren
basis~\cite{vanNeerven:1983vr}. We begin by parametrizing the loop
momentum of each strand in the given diagram $\ell_l$ (with $l=1,2$
for all factorized (one-loop squared) diagrams and $l=1,2,3$
otherwise)
as~\cite{Ita:2015tya,Abreu:2017xsl,Abreu:2017hqn}:
\begin{align} \begin{split}
    \ell_l &=
        \sum_{j \in B^p_l} v^j_l r^{l j} + \sum_{j \in B^t_l} v^j_l \alpha^{l
        j} + \sum_{i \in B^{ct}} \frac{n^i}{(n^i)^2}  \alpha^{l i} + \sum_{i \in B^{\epsilon}}
        n^i  \mu_l^{i}, \label{eq:AdaptedCoordinates} \\ 
        r^{l j}& = -\frac{1}{2} ( \rho_{lj} - (q_{lj})^2 - \rho_{l(j-1)} + (q_{l(j-1)})^2 )\,,  \\
        \mu_{ll}&\equiv \mu_l\cdot\mu_l 
        = \rho_{l0} - \sum_{\nu = 0}^3\ell_l^\nu \ell_{l\,\nu}\,,
        \end{split}
\end{align} 
where the index sets $B_l^p$, $B_l^t$, $B^{ct}$ and $B^\epsilon$ have the
following physical interpretation: $B_l^p$ 
is the \emph{physical}
space spanned by
the momenta attached to the strand $l$; $B^{ct}$ is the physical (4-$D$)
\emph{common transverse} space (transverse to all $B_l^p$'s); 
$B_l^{t}$ is the
\emph{transverse} space to $B_l^{p}$ not overlapping $B^{ct}$; and $B^\epsilon$ is
the extra dimensional $D-4$ space. The $q_{lj}$ are linear combinations of
external momenta. The vectors $v^j_l$ are a basis of the 
physical scattering plane chosen to be dual to the external
momenta, and the $n^i$ are an orthogonal basis of the space
transverse to the physical scattering space.

In this coordinate system, it is trivial to satisfy the on-shell
conditions by setting the associated propagator variables $\rho$ to
0. Further, it is clear that for any finite field valued choice
of the
irreducible scalar products $\alpha$, the 4-dimensional components of
the loop momenta also belong to the finite field. In order to treat
the typically algebraic valued $(D-4)$ dimensional components of the
loop momenta, $\mu_i$, we represent them abstractly, taking the
$\mu_i$ themselves as basis vectors,
\begin{equation}
  \ell_l^{(D-4)} = w_{l,1} \mu_1 + w_{l,2} \mu_2.
\end{equation}
This choice must then be respected by the scalar product, that is
\begin{equation}
  \ell_r \cdot \ell_s = \ell^4_r \cdot \ell^4_s - \sum_{i,j = 1}^{2} w_i^r w_j^s \mu_{ij}.
\end{equation}

A further complexity when performing unitarity-based computation in
finite fields, is the handling of the internal states that appear on a
cut propagator. In these states, normalization factors appear
that cannot be represented in the finite field. Instead of state
sums, we insert a light-cone projector operator in the
$D_s$-dimensional space of the polarization vectors
\begin{equation} P_l^{\mu \nu} = -g^{\mu \nu} + \frac{\ell_l^\mu \eta^\nu +
    \eta^\mu \ell_l^\nu}{\eta\cdot \ell_l}\ , \label{eq:LightConeProjector}
\end{equation} 
where $\eta$ is an arbitrary light-like reference vector with rational
components (which can of course be represented in the finite field) and that
must fulfil $\eta \cdot \ell_l \ne 0$ and
$\eta\cdot \mu_i = 0$. Although
this is a sound solution, we have also implemented
modified physical states for performing the calculations as it
is more efficient for larger values of $D_s$.
In this way, intermediate square roots can be avoided in a gluonic
numerical unitarity calculation. More details are given
in~\cite{Abreu:2017hqn}.

\section{The planar two-loop five-gluon helicity amplitudes}\label{sec:5g}
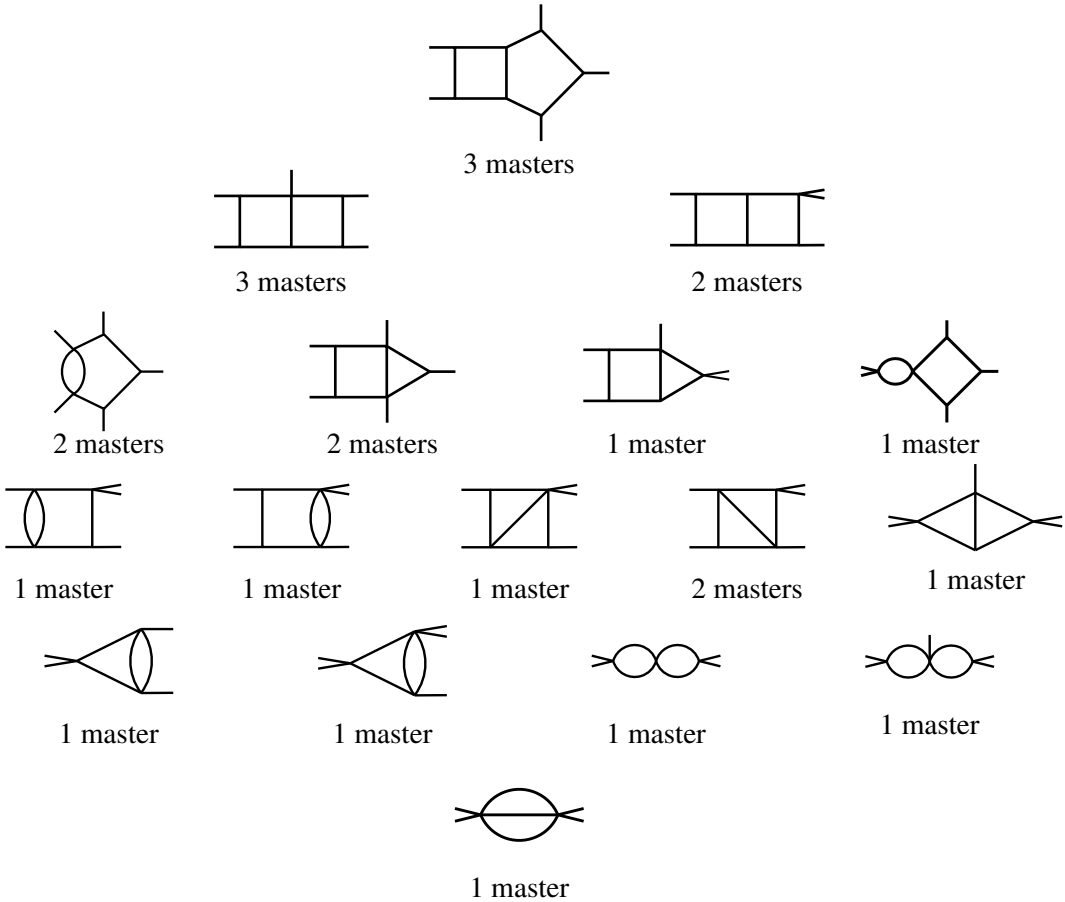
\begin{figure}[!h] \begin{tikzpicture}[scale=1.2]
    \node at (5,0){\includegraphics[scale=0.4]{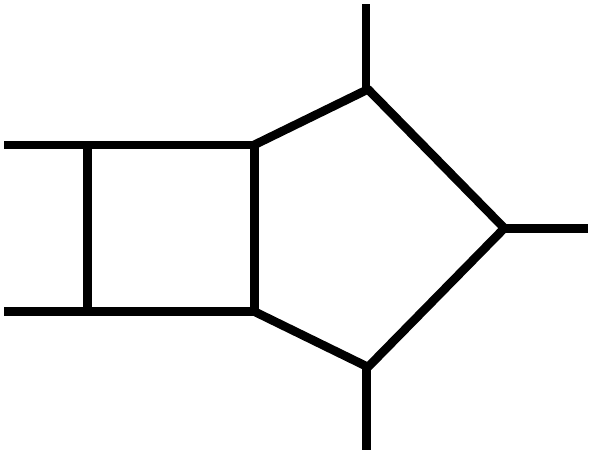}};
    \node at (5,-1){3 masters};
    \node at
    (2.5,-1.5){\includegraphics[scale=0.4]{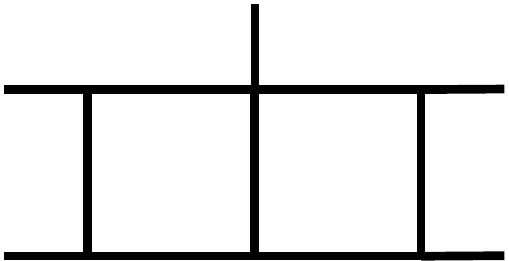}};
    \node at (2.5,-2.3){3 masters};
    \node at
    (7.5,-1.6){\includegraphics[scale=0.4]{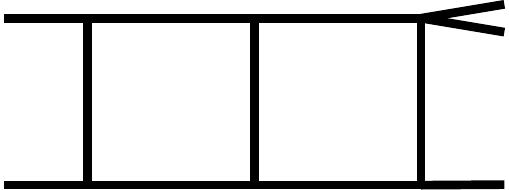}};
    \node at (7.5,-2.3){2 masters};
    \node at
    (0.5,-3.3){\includegraphics[scale=0.35]{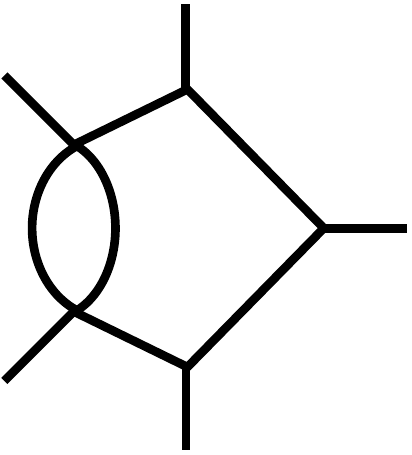}};
    \node at (0.5,-4.1){2 masters};
    \node at
    (3.5,-3.3){\includegraphics[scale=0.4]{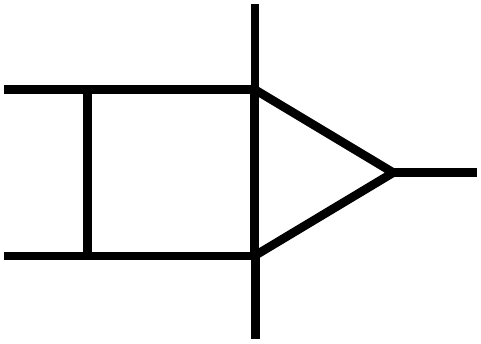}};
    \node at (3.5,-4.1){2 masters};
    \node at
    (6.5,-3.2){\includegraphics[scale=0.4]{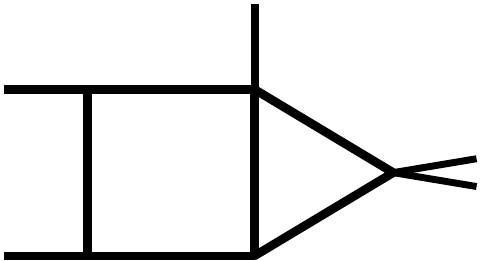}};
    \node at (6.5,-4.1){1 master};
    \node at
    (9.5,-3.3){\includegraphics[scale=0.4]{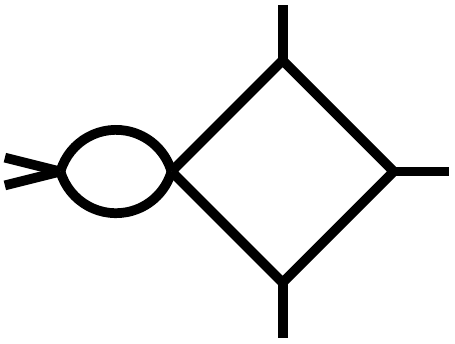}};
    \node at (9.5,-4.1){1 master};
    \node at
    (0,-4.9){\includegraphics[scale=0.45]{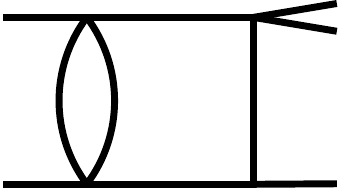}};
    \node at (0,-5.7){1 master};
    \node at
    (2.5,-4.9){\includegraphics[scale=0.45]{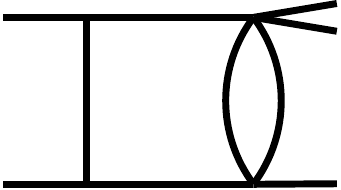}};
    \node at (2.5,-5.7){1 master};
    \node at
    (5,-4.9){\includegraphics[scale=0.45]{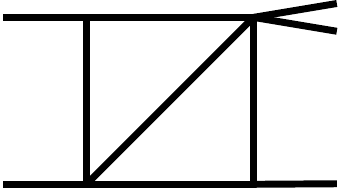}};
    \node at (5,-5.7){1 master};
    \node at
    (7.5,-4.9){\includegraphics[scale=0.45]{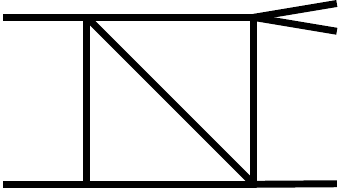}};
    \node at (7.5,-5.7){2 masters};
    \node at
    (10,-4.8){\includegraphics[scale=0.45]{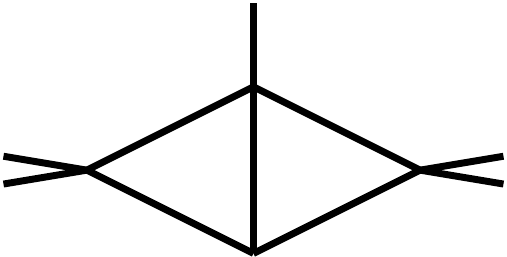}};
    \node at (10,-5.6){1 master};
    \node at
    (0.5,-6.5){\includegraphics[scale=0.5]{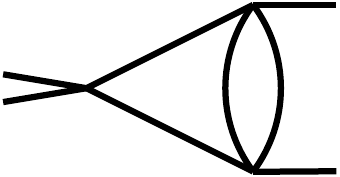}};
    \node at (0.5,-7.3){1 master};
    \node at
    (3.5,-6.5){\includegraphics[scale=0.5]{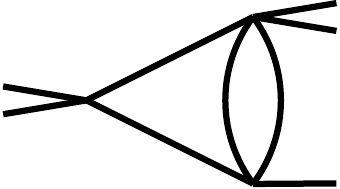}};
    \node at (3.5,-7.3){1 master};
    \node at
    (6.5,-6.5){\includegraphics[scale=0.5]{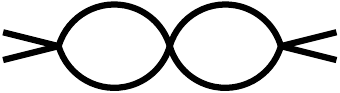}};
    \node at (6.5,-7.3){1 master};
    \node at
    (9.5,-6.45){\includegraphics[scale=0.5]{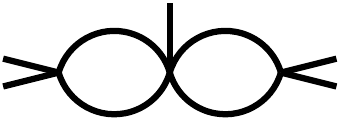}};
    \node at (9.5,-7.2){1 master};
    \node at
    (5,-8.2){\includegraphics[scale=0.6]{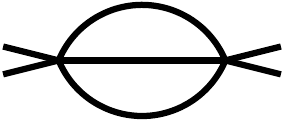}};
    \node at (5,-9){1 master};
\end{tikzpicture} \caption{Propagator structures with master integrals for
massless two-loop five-point amplitudes.}
\label{fig_master_int}
\end{figure}

The planar diagram hierarchy for two-loop five-point massless
amplitudes is comprised of 60 topologically independent diagrams, in
which 25 master integrals appear. In figure~\ref{fig_master_int} we
show the subset of diagrams with master integrals and their
corresponding
multiplicity. In total, considering all permutations of the external
legs maintaining the cyclic order, we need to compute 155 integral
coefficients.

We present results for the four independent helicity configurations
(up to cyclic permutations and parity transformations) that appear for
the scattering of five gluons. We include all-plus and
single-minus helicity configurations, though they are not
necessary
for a potential NNLO QCD calculation of three-jet production, as their
corresponding tree-level amplitudes vanish. The results we
present
for the amplitudes are obtained in the Euclidean phase-space
point
\begin{align}\begin{split}
  p_1 &= \left(\frac{1}{2}, \frac{1}{16}, \frac{i}{16}, \frac{1}{2}\right), \quad
  p_2 = \left( -\frac{1}{2}, 0, 0, \frac{1}{2} \right),\quad
  p_3 = \left(\frac{9}{2}, -\frac{9}{2}, \frac{7i}{2}, \frac{7}{2}\right), \\
  p_4 &= \left(-\frac{23}{4}, \frac{61}{16}, -\frac{131i}{16}, -\frac{37}{4}\right), \quad
  p_5 = \left(\frac{5}{4}, \frac{5}{8}, \frac{37i}{8}, \frac{19}{4}\right), 
  \label{eq:EvalPoint}
\end{split}\end{align}
with the corresponding invariants
 $s_{12} = -1,\  s_{23} =
-8,\ s_{34} = -10,\
s_{45} = -7,\  s_{51} = -3$. We set the renormalization scale
$\mu$ to
$\mu=1$.
We present results in the 't Hooft-Veltman scheme for dimensional
regularization~\cite{tHooft:1972tcz}. As our coefficients have been computed in
an exact manner, the numerical precision of our results is only
limited by the
precision to which the analytic expressions of the integrals  are computed~\cite{Papadopoulos:2015jft,Gehrmann:2000zt}. For the
evaluation of the generalized polylogarithms we employ
\texttt{GiNaC}~\cite{Vollinga:2004sn}.

\begin{table}[htb]
\hspace{-3mm}
\centering
\begin{tabular}{|c|c|c|c|c|c|}
  \hline
  $\mathcal{A}^{(2)}/\mathcal{A}^{({\rm norm})}$ & $\epsilon^{-4}$ & $\epsilon^{-3}$ & $\epsilon^{-2}$ & $\epsilon^{-1}$ & $\epsilon^{0}$ \\ \hline
   $(1^+, 2^+, 3^+, 4^+, 5^+) $  & --- & --- & -5.0000000  &   -3.89317903  & 5.98108858  \\ \hline
   $(1^-, 2^+, 3^+, 4^+, 5^+) $  & --- & --- & -5.0000000  &  -16.3220021  &  -10.3838132   \\ \hline 
  $ (1^-, 2^-, 3^+, 4^+, 5^+) $ &  12.50000 &  25.462469  &  -1152.8431  &  -4072.9383  &  -3637.2496  \\ \hline
  $ (1^-, 2^+, 3^-, 4^+, 5^+) $ & 12.50000  &  25.462469  &  -6.1216296  &  -90.221842  &  -115.78367  \\ 
  \hline
\end{tabular}
\caption{Numerical results for all types of helicity configurations of the
two-loop five-gluon amplitudes. For the amplitudes with vanishing trees, we set
$\mathcal{A}^{({\rm norm})}=\mathcal{A}^{(1)}(\epsilon=0)$, and for the MHV
configurations $\mathcal{A}^{({\rm norm})}=\mathcal{A}^{(0)}$. Details are discussed
in the text.}
\label{tab:Results}
\end{table}

In table~\ref{tab:Results} we show our results for the helicity amplitudes. The
results for amplitudes that vanish at tree-level have
been normalized to
$\mathcal{A}^{({\rm norm})}=\mathcal{A}^{(1)}(\epsilon=0)$, that is to the
corresponding (finite) one-loop amplitudes at order 
$\epsilon^0$. The other
amplitudes are normalized to the corresponding tree-level
amplitudes. By this
choice of normalization we remove any unphysical phase and expose more clearly the
structure of the infrared poles. All infrared poles have been
successfully checked against
the expected results~\cite{Catani:1998bh}.
Furthermore, we
have cross-checked against the known all-plus five-point
results~\cite{Badger:2013gxa,Badger:2015lda,Gehrmann:2015bfy,Dunbar:2016aux}
as well as validated the results of~\cite{Badger:2017jhb}.

Obtaining the results in table~\ref{tab:Results} requires a large
number
of operations, including the sampling of a large number of trees and
of the integrand function bases. This is performed over ten values of
$D$ and three values of $D_s$ to reconstruct the dependence on the
regulators. Although we have implemented structure for information
caching, this nevertheless amounts to an under three-minute calculation for
the most complex helicity structure in a single finite field with
a
single-threaded calculation performed on an \texttt{i7} Intel
processor. We expect improvements in the efficiency of the calculation
in the future, and we have also explored several parallelization
strategies that can scale considerably computation times.

\section{Outlook}\label{sec:outlook}
We have presented the first calculation of a physically-relevant two-loop
five-point amplitude with a full reduction to master integrals. The computation
of the integral coefficients was achieved with the recently developed multi-loop
numerical unitarity method, which exploits an integrand decomposition based on
master integrands and surface terms.
Importantly, it now feasible to
assemble our amplitudes into squared matrix elements which can then be
integrated over the physical phase space.

It is well known that unitarity-based techniques have good scaling properties as
a function of the number of external legs (and also of the number
of loops) making them very suitable
for studying generic scattering amplitudes in field theories. While computing our results for four-point
and five-point amplitudes, we observed that even at this early stage
of the new developments, the computation of six-point two-loop amplitudes is
within reach.  Of course, these cases also pose the challenge of
computing the corresponding master integrals, but we expect this to be
possible in the near future.

Although gluon amplitudes are an excellent playground, we look
forward to apply our techniques to more generic processes that can
have interesting phenomenological applications.  This includes
processes with quarks, leptons and electro-weak bosons. Adding additional
scales, for example by having external massive particles, poses only
minor problems.

We have for the first time implemented an approach to reduce a
full amplitude to master
integrals employing finite field arithmetics. Such an approach holds
several advantages, including the possibility of obtaining exact
results for the integral coefficients when using rational 
phase-space points and exploiting integer-based operations for which
modern CPUs are extremely efficient.  For future applications to
phenomenological calculations, we envisage a hybrid approach that
makes use of both exact and floating point operations.  

Another potential application of our exact integral coefficients is
the exploration of the analytical structure of the
amplitudes. Techniques for multivariate functional reconstruction have
recently generated a lot of interest~\cite{Peraro:2016wsq}, which,
when combined with fast amplitude evaluations, open a path towards
obtaining analytic results for the amplitudes. 

We have developed a computational framework for multi-loop numerical unitarity
which we plan to make public to the wider high-energy physics community in 
the near future.
We are planning to release the implementation in
phases, starting with a version that gives access to the amplitudes
that we have computed. Over time, we hope to provide more generic
two-loop matrix elements that can be used for
phenomenologically-relevant NNLO QCD computations.

\acknowledgments{
We thank the organizers for the opportunity to present our work at
this conference.
The work of S.A., F.F.C., and B.P. is supported by the Alexander von
Humboldt Foundation, in the framework of the Sofja Kovalevskaja Award
2014, endowed by the German Federal Ministry of Education and
Research.
The work of M.Z. is supported by the U.S. Department of Energy under
Award Number DE-{S}C0009937.
}
\bibliographystyle{JHEP}
\bibliography{NumUni2_LL2018}

\end{document}